\documentstyle[amsbsy,epsfig]{espart}
\begin{document}
\def\be{\begin{equation}}
\def\ee{\end{equation}}
\def\bq{\begin{eqnarray}}
\def\eq{\end{eqnarray}}
\def\d{\mbox{\rm d}}
\def\bra#1{\langle #1 \,\vert}
\def\ket#1{\vert\, #1 \rangle}
\def\braket#1#2{\langle #1 \,\vert\, #2 \rangle}

\def\newpage{\vfill\eject}
\def\newline{\hfill\break}

\begin{frontmatter}
\title{Recoil effects on nucleon electromagnetic form factors}

\author{Alessandro Drago$^{a}$, Manuel Fiolhais$^{b}$ and Ubaldo Tambini${^a}$}
\address{$^{a}$Dipartimento di Fisica, Universit\`a di Ferrara,
and INFN, Sezione di Ferrara,\\
Via Paradiso 12, Ferrara, Italy 44100\\
$^{b}$Departamento de F{\'\i}sica da Universidade \\
and Centro de F{\'\i}sica Te{\'o}rica, P-3000 Coimbra, Portugal}
\date{\today}
%
%
\begin{abstract}
The electromagnetic form
factors are computed
using eigenstates of linear momentum for the nucleon. The latter is described
in the framework of the chiral color-dielectric model, projecting
the hedgehog ansatz on eigenstates of angular momentum and isospin.
Form factors are well reproduced, with the exception of the magnetic one
for the proton, up to $q^2 \simeq 0.5$ GeV$^2$.
The effect of the
removal of the spurious center-of-mass contributions shows up
mainly in the electric form factor of the proton. A noticeable
improvement is obtained with respect to the calculation without
linear momentum projection.

\end{abstract}
\end{frontmatter}
\section{Introduction}

We report on a theoretical calculation of the electromagnetic form-factors
of the nucleon in the space-like region, performed
in the framework of
an effective model --- the chiral color-dielectric
model (CDM) \cite{pirner,birserev} ---  in which  the
nucleon is described as a chiral soliton. The model
contains quark and meson degrees of freedom
and a phenomenological scalar field which is responsible for quark
confinement.

In the previous calculations of the form factors in the framework
of the chiral soliton models of the nucleon,
as the linear sigma model \cite{sigma}, the Nambu-Jona-Lasinio
model \cite{njl}, the Skyrme model \cite{skyrme} and the
CDM \cite{FN94},
it was always assumed
that the
nucleon is at rest before and after the interaction with the virtual photon,
the so-called static approximation.
In the present work we overcome, at least in part, the technical
difficulties associated with 
the computation of the form factors when
the nucleon initial and final states
are eigenfunctions of  linear momentum, at least non relativistically.
Our formalism  is a generalization  of the one presented in
a  work by Neuber {\em et al.} \cite{NF93,Neuber},
where {\em static} properties of the nucleon have been
computed in the framework of the CDM.
In their case therefore it was enough to build eigenstates of the angular
momentum  having {\em zero} linear momentum.

The main technical problem in our calculation is due to the
non-commutativity
of the projectors on linear and on angular momentum. It will be
shown that the problem can be solved by taking a suitable 
average on the direction
of the transfered momentum,
as it is suggested by the equation
$\int{\d}\hat{\vec q} \,[P_{\vec q},P_{JM}]=0$.
Morover it will be shown that the 
Fourier transform of the matrix elements of the electromagnetic current 
does not
depend on the direction of the momentum transfered, 
if nucleon states are considered. Therefore the
integration on $\hat{\vec q}$
does not imply any approximation. 
The use of the {\it effective commutativity} of the two projectors
simplifies the computation of the doubly projected form factors.
These mathematical aspects are applicable in general
to any quark-meson chiral soliton model.

The numerical results shown  in this paper refer to the electric and
the magnetic proton and neutron form factors computed in a particular version
of the chiral CDM. This is the so-called ``single minimum" version
which gives good results in
quark matter calculations \cite{DFT95}. 
The model contains two parameters that we adjust in order to reproduce
the average $\Delta$--$N$ mass  and the isoscalar nucleon radius.
All other results are obtained without any further parameters' fitting.

This paper is organized as follows. In Section 2 the electromagnetic
form-factors are defined. In Section 3
we review the chiral color-dielectric model and the way
model states representing a nucleon with definite momentum are obtained.
Section 4  is devoted to the  formalism to compute
the electric and the magnetic form factors
of the nucleon in the projected
hedgehog state. Finally, in Section 5, the results are presented and
discussed.

\section{The electromagnetic form factors of the nucleon}

Let $|N_\alpha(p_{\mu})\rangle$ represent a nucleon state of mass $M_N$, with
spin and isospin
described by $\alpha$.  The standard definition of the
electromagnetic form factors
is given by:
\be
\bra{N_f(p'_{\mu})} J^\mu_{em}(0)\ket{N_i(p_{\mu})}
=\bar u_f(p'_{\mu})\left\lbrack F_1(q_{\rho}^2)\gamma^\mu+
i\,{F_2(q_{\rho}^2)\over 2M_N}\sigma^{\mu\nu}q_\nu\right\rbrack u_i(p_{\mu}).
\label{21}
\ee
They only depend on the modulus of the
momentum transfer of the virtual photon $q_{\mu}=p'_{\mu}-p_{\mu}$,
being real functions  of  $q_{\mu}q^{\mu}=(q^0)^2- {\vec
q}^2$.
We work in the Breit frame
where the photon 4-momentum is $q^{\mu} =(0,{\vec q})$, i.e.
the energy transfer is zero. Our results will be presented
as a function of $q=\vert {\vec q} \vert$.

In the Breit frame, Eq.  (\ref{21}) reads explicitly
\be
\bra{ N_f ({ {\vec q}\over 2})   }  {J}^{\mu}_{em}(0) \ket{
N_i (-{ {\vec q}\over 2}) }
= \overline{u}_f({ {\vec q}\over 2} )\left[ F_1(q^2) \gamma^{\mu} +
i \frac{F_2(q^2)}{2 M_N}\sigma^{\mu \nu} q_{\nu}
\right]u_i(-{ {\vec q}\over 2}).
                   \label{25}
\ee
Normalizing the Dirac spinors so that  $\overline{u} u =1$,
\be
u({\vec p})= \sqrt{\frac{E+M_N}{2 M_N}} \left(
\begin{array}{c}
1 \\
\frac{{{\vec \sigma} \cdot {\vec p}}}{E+M_N}
\end{array}\right)\ket{\chi} ,
                    \label{26}
\ee
the matrix element  (\ref{25}) can be worked out yielding
\bq
\bra{ N_f ({ {{\vec q}}\over 2})}  &{J}^{\mu}_{em}&(0) \ket{
N_i (-{ {{\vec q}}\over 2}) } \nonumber \\
&=& F_1(q^2) \left[ \bra{\chi_f} \chi_i \rangle \delta_{\mu 0} + i
\sum_{j=1}^3 \frac{\bra{\chi_f} [ {{\vec \sigma} \times {\vec q}}
]_j \ket{\chi_i}}{2 M_N}
\delta_{\mu j} \right]  \nonumber \\
&-&F_2(q^2) \left[ \bra{\chi_f} \chi_i \rangle
\frac{q^2}{4 M_N^2}\delta_{\mu 0}
 - i \sum_{j=1}^3 \frac{\bra{\chi_f} [ {{\vec \sigma} \times {\vec q}} ]_j
\ket{\chi_i}}{2 M_N}
\delta_{\mu j} \right].
\label{27}
\eq

From $F_1(q^2)$ and $F_2(q^2)$ it is usual to define the so-called Sachs
form factors, $G_E(q^2)$ and $G_M(q^2)$, where {\small $E$} and
{\small $M$} stand for
``electric" and ``magnetic" respectively, which are expressed by
\bq
G_E(q^2)&=&F_1(q^2)-{q^2\over 4M_N^2}\ F_2(q^2) \label{28}\\
G_M(q^2)&=&F_1(q^2)+ F_2(q^2). \label{29}
\eq
Using these definitions and Eq. (\ref{27}) one obtains explicit
formulas for the electric and the magnetic form factors:
\be
G_E(q^2) \bra{\chi_f} \chi_i \rangle =
\bra{N_f ({ {\vec q}\over 2})}  {J}^{0}_{em}(0) \ket{
N_i (- { {\vec q}\over 2} ) }  \label{210}
\ee
\be
i {G_M(q^2) \over 2 M_N} \bra{\chi_f} {\vec \sigma \times \vec q}
\ket{\chi_i} =
\bra{N_f ({ {\vec q}\over 2})}  {{\vec J}}_{em}(0) \ket{
N_i (-{ {\vec q}\over 2} ) }\, .  \label{211}
\ee

\section{The projected  chiral color-dielectric model}

In this work we shall use a  chiral version of the CDM,  whose
Lagrangian reads
\cite{pirner,birserev,NF93}
\bq
{\cal L} &=& i\bar q\gamma^{\mu}\partial_{\mu}q
       +{g\over \chi} \bar q\left(\sigma_o
+i\gamma_5\vec\tau\cdot\vec\pi\right) q
                       +{1\over 2}{\left(\partial_\mu\chi\right)}^2
                       -V\left(\chi\right) \nonumber \\
        &+&{1\over 2}{\left(\partial_\mu\sigma_o \right)}^2
                       +{1\over 2}{\left(\partial_\mu\vec\pi\right)}^2
                       -U\left(\sigma_o ,\vec\pi\right).
                                                           \label{31}
\eq
This is a model with interacting quarks, chiral mesons $\sigma_o$ and
 $\vec{\pi}$
and also a chiral singlet scalar field $\chi$ responsible for
 confinement. The potential
$U\left(\sigma_o ,\vec\pi\right)$ in  ({\ref{31}}) for the chiral mesons
is the so-called `mexican-hat' potential
\be
U\left(\sigma_o ,\vec\pi\right)={\displaystyle {{\lambda^2} \over 4}}
\left(\sigma_o^2+\vec\pi^2-\nu^2\right)^2+c \sigma_o+d. \label{32}
\ee
The chiral symmetry
$SU(2)\times SU(2)$ of ${\cal L}$ is
explicitly broken by the small term $c \sigma_o$ in (\ref{32});
the last term in the same expression is a constant fixed in order to have
${\mathrm{min}} \,\, U = 0$.
The parameters $\lambda$ and $\nu$ are related to the
sigma and the pion masses and to the pion decay constant:
\be
\lambda^2= \frac{m_{\sigma}^2- m_{\pi}^2}{2 f_{\pi}} ; \qquad
\nu^2 =f_{\pi}^2 - \frac{m_{\pi}^2}{\lambda^2}.
\ee
For the $\chi$ field potential we consider  the quadratic form
\be
V(\chi)=\frac12 M^2 \chi^2\, , \label{33}
\ee
where $M$ is the $\chi$ mass.
 It is well known that the chiral CDM allows for soliton
solutions in which the quarks are absolutely confined
\cite{pirner,birserev,NF93}. In such
solutions the $\chi$ mean field is a decreasing
function of the distance, approaching zero in the limit $r \rightarrow \infty$.
This generates a raising dynamical mass for the quarks and
confines them.
In previous works an
exhaustive study of the
model with a quartic  (or `double minimum') potential was carried out
\cite{FN94,NF93}.
In this work we consider just the  quadratic (`single minimum') potential for
the confining field.
We recently  showed \cite{DFT95}  that for `double
minimum' potentials
and for all sets of parameters fitting nucleon properties, the
equation of state for quark matter turns out to be unrealistic. Indeed
even at very low density the energy {\it per} baryon number for quark matter
turns out to be smaller than that for nuclear matter. Using instead
a quadratic potential for the $\chi$ field
a realistic equation of state is obtained.

Altogether, the parameters of the model defined by (\ref{31})
 are the pion and sigma masses
(fixed at $m_\pi=0.139$ GeV and $m_\sigma$=1.2 GeV), the pion decay constant
($f_\pi=0.093$ GeV), and $g$ and $M$,
 the quark-meson-$\chi$
coupling constant and the $\chi$-mass, respectively.

In order to obtain model states representing the nucleon we used the procedure
explained in great detail in Ref. \cite{NF93} which, for the
sake of completeness, is sketched here. We consider three valence quarks in
the hedgehog state $\ket{h}=
\frac{1}{\sqrt{2}} \left(\vert u\downarrow\,\rangle-
\vert d\uparrow\,\rangle \right) $,
all occupying the same lowest positive energy s-orbital and
surrounded by clouds
of $\chi$, sigmas and pions, described by coherent
states ( $\ket{\Pi}$, for pions; $\ket{\Sigma}$, for sigmas; and $\ket{\chi}$,
for the confining field).
The meson mean fields are the expectation values of the field operators
in the corresponding coherent states.

We can write the quark single particle states and the meson mean fields
as
\bq
& &
\langle\,{\vec r} \vert q\,\rangle = {1\over \sqrt{4\pi}}
\pmatrix{u(r) \cr \imath v(r){\vec\sigma}\cdot\hat{\vec r} \cr}
\ket{h}  \label{34}   \\
& & \sigma_o ({\vec r}) = \bra{\Sigma} {\sigma} \ket{\Sigma} - f_\pi
=\sigma (r) -f_\pi \label{35}    \\
& & \vec\pi ({\vec r}) = \bra{\Pi} {\vec\pi} \ket{\Pi}
= {{\vec r}\over r}\phi(r)  \label{36} \\
& & \chi ({\vec r}) = \bra{\chi} {\chi} \ket{\chi}
= \chi (r)\, , \label{37}
\eq
where ${\sigma}$ is just the fluctuating part of the $\sigma_o$ field.
Altogether, the hedgehog ansatz reads
\be
\vert \psi_{hh}\,\rangle =
\left( C_{h}^{\dag}\right)^3
\ket{\Pi} \, \ket{\Sigma} \, \ket{\chi}, \label{38}
\ee
where $C_h^{\dag}$ creates a particle in the single quark state
(\ref{34}).

Of course, solitons described
by the hedgehog $\ket{\psi_{hh}}$ cannot represent
physical baryons because they are not eigenstates
of angular momentum or isospin. In addition, (\ref{38}) represents a localized
object and therefore the translational symmetry of the model hamiltonian is
also broken in such states. In particular they contain
spurious centre-of-mass components which contribute to the energy and to
the other observables. However, a nucleon at rest can be obtained
by applying the projector
onto linear momentum ${\vec q}=0$
together with the projector onto  angular momentum-isospin.
The linear momentum projector is given by \cite{RS}
\be
P_{{\vec q}} = \left({1\over 2\pi}\right)^3\,
\int {\d}\vec a\,
{\rm e}^{\imath{\vec a}\cdot{\vec q}}U({\vec a}), \label{39}
\ee
where $U({\vec a})$ is the translation operator.
It is well known that due to
the symmetry of the hedgehog it is enough to
perform a single projection (e.g. onto spin), since this automatically projects
onto the same value of isospin.
The operator which projects out from the hedgehog  a state with
angular momentum $J$ and isospin $T=J$  is
\be
P_{MM_T}^J=(-1)^{J+M_T} {2J+1\over 8\pi^2}
\int \d^3\Omega\,{\cal D}^{J\,^ *}_{M,-M_T}(\Omega )R(\Omega ),
\label{310}
\ee
where
$\Omega =(\alpha ,\beta ,\gamma )$ are the three Euler angles,
${\cal D}^J_{M,K}(\Omega )$ are the
Wigner functions and $R(\Omega )$ is the rotation operator. In the
following we will
consider $M_T=-M$
and  use the shorthand notation $P_{JM} \equiv P_{M \,-M }^J$.

The radial functions
in (\ref{34})-(\ref{37}) are determined using an approximate
variation-after-projection method firstly suggested by Leech
and Birse
\cite{birsecm}. They are computed by
minimizing the
expectation value of the (normal-ordered) model hamiltonian in the
model baryon state with quantum numbers $J=T=\frac12$
and linear momentum zero:
\be
P_{{\vec q}=0}P_{JM}\,\vert\Psi_{hh}\,\rangle =
\vert J,T,M,{\vec q}=0\,\rangle \, . \label{311}
\ee

In the model there are two parameters, $g$ and $M$, yet to be fixed.
However, because of the smoothness of the $\chi$-field in a typical soliton
solution and of the relative weakness of the chiral meson clouds, the relevant
parameter turns out to be $G=\sqrt{g M}$. In the quark matter sector this is
indeed the only free parameter of the model \cite{Judith}.
If $G$ is fixed to reproduce the isoscalar radius of the nucleon
one obtains $G=0.2~$ GeV. For this parameter the nucleon-delta average mass
is around 1.13 GeV (experimental value 1.085 GeV). It is interesting to
observe that if $g$ and $M$ are changed, keeping $G$ fixed, the static
properties of the nucleon are essentially unchanged.
For example, for $G=0.2$ GeV, changing the mass of the $\chi$ field in the
range 0.8--2.0 GeV, affects the results by less than 1\%.

In the present version of the CDM, the
nucleon-delta mass splitting results only from the quark-pion
interaction. Due to the weakness of the pionic field,
the nucleon delta mass splitting obtained is too small.
The experimental value of the splitting could be recovered
if, in addition,
a  color-magnetic
interaction (like in the MIT bag model or in the cloudy bag model) was
considered \cite{GS93}. We will come back to this point in the conclusions.

\section{Electromagnetic form factors in the projected hedgehog}

In order to compute
the electromagnetic nucleon form factors one has to evaluate
matrix elements of the electromagnetic current operator. In the CDM
the latter is given by
\be
 J^\mu_{em} (x) =  :  \sum_{a=1}^3\, \bar q_a(x)\gamma^{\mu}
\left({1\over 6}+{\tau_3\over 2}\right)q_a(x)+
[\vec{\pi}(x)\wedge\partial^{\mu}\vec{\pi}(x)]_3 \, :  .\label{312}
\ee

As mentioned in the Introduction, in the past
the matrix elements of this operator  have been
computed in the {\em static} approximation (the
nucleon is assumed to be at rest {\em before} and {\em after} the
interaction with the photon).
In the present
work we go beyond this approximation, since we compute the matrix
elements using momentum eigenstates. In principle these should be obtained
by boosting \cite{Lu} the nucleon zero momentum eigenstate (\ref{311}).
However, the technical difficulties associated
with  boosting are prohibitive. We
approximate this operation by a Peierls-Yoccoz projection,
i.e. we consider our model state with momentum ${\vec q}$
to be given by

\be
\vert N ({q})\,\rangle\to \sqrt{(2\pi)^3 \delta^ 3(0)}
\sqrt{\frac{E}{M_N}}
\frac{P_{{\vec q}}\,\vert\Psi_{JM}\,\rangle}
{\sqrt{ \langle \,P_{{\vec q}}\,\Psi_{JM}\,
\vert P_{{\vec q}}\,\Psi_{JM}\,\rangle}}
\, , \label{313}
\ee
where the square roots are just normalization factors and
\be
\vert\Psi_{JM}\,\rangle = P_{JM}\,\vert\Psi_{hh}\,\rangle.
\ee
The approximation involved in assuming projected
instead of boosted states is valid for small $\vec q$.

Before presenting the formalism to compute the form factors
as matrix elements of
the electromagnetic current (\ref{312}) taken between nucleon states,
let us recall that
\be
\int {\d}{\vec z}\,  \left [F({\vec z}) U({\vec z}),R({\Omega})\right
]=0,
\label{theo}
\ee
if $F({\vec z})$ is a scalar function of ${\vec z}$ \cite{NG92}.
This can be seen writing explicitly the commutator, rotating the
argument of $F$ and $U$ and redefining
${\vec z'}={\cal R}({\Omega}){\vec z}$.
The previous commutation relation will be very useful in the
following.

Another important point is that, due to the symmetry of the hedgehog,
rotations of this state in spin or isospin space
are equivalent. Therefore,  as it was
already pointed out in the previous Section, projecting the hedgehog
onto spin $J$ implies a simultaneous projection onto
isospin $T=J$, and the two projections are equivalent. Hence,
the following relations hold:
\be
P_{\vec q} P_{JM}\vert \Psi_{hh}>=
P_{\vec q} P_{TM}\vert \Psi_{hh}>=
P_{TM} P_{\vec q}\vert \Psi_{hh}>\ne
P_{JM} P_{\vec q}\vert \Psi_{hh}>, \label{commut}
\ee
where the commutation between operators working in isospin space
and operators working in ordinary space has been used. The projector
$P_{TM}$ is defined similarly to the projector $P_{JM}$ (Eq.
(\ref{310})),
but with the rotation operator $R$ acting in isospin space, replacing the
rotation operator in spin space. We shall exploit
relations (\ref{commut}) in the evaluation of the
magnetic form factors.

\subsection{Electric form factor}

From the definition of the electric form factor (\ref{210}) and
using the correspondence (\ref{313}), the electric form-factor is given
by
\be
G_E(q^2)=\frac{E}{M_N}
\frac{\int {\d}{\vec x} \, \langle P_{\frac{{\vec q}}{2}} \Psi_{JM}
\vert J_{em}^0(0) \vert P_{-\frac{{\vec q}}{2}} \Psi_{JM}\rangle}
{\langle P_{\frac{{\vec q}}{2}} \Psi_{JM} \vert
P_{\frac{{\vec q}}{2}} \Psi_{JM} \rangle} \, ,
\ee
and using the explicit form of the linear momentum projector (\ref{39})
one obtains
\be
G_E(q^2)=\frac{E}{M_N}
\frac{\int {\d}\vec x \, {\d}\vec b\,  {\d}\vec b' \,
e^{-i ({\vec b}+{\vec b'})
\cdot \frac{{\vec q}}{2}}  \,
 \langle U({\vec b'}) \Psi_{JM} \vert
J_{em}^0(0) \vert U({\vec b}) \Psi_{JM}\rangle}
{ \int {\d}\vec b \, {\d}\vec b' \, e^{-i ({\vec b}-{\vec b'})\cdot
\frac{{\vec q}}{2}} \, \langle U({\vec b'}) \Psi_{JM} \vert
 U({\vec b}) \Psi_{JM} \rangle} \, .
\ee

In principle the form factors should be functions of $q^2$ only.
However, due to the approximate treatment of the center-of-mass motion
this is in general 
no longer guaranteed and there is a spurious dependence on the
angle between $\bf q$ and the quantization direction.
However, if $J=1/2$ states are considered, it is possible to
show that the form factor is indeed a function of $q^2$ only. 
In fact, due to parity, the Fourier transform
of the matrix element of the current has to be a function
of $(\bf q\cdot\bf J)^2$ which, for $J=1/2$, is proportional
to $q^2$.
We can therefore integrate the 
direction $\hat{\vec  q}$, both in the current matrix element
and in the normalization factor at the denominator.
After the expansion of the exponentials in spherical waves,
only the $\ell =0$ wave contributes and one gets
\be
G_E(q^2)=\frac{E}{M_N}
\frac{\int {\d}\vec x\, {\d}\vec b \, {\d}\vec b' \,
j_0\,\left( \vert \frac{{{\vec b}}+{{\vec b'}}}{2} \vert  q \right)
 \langle U({\vec b'}) \Psi_{JM} \vert
J_{em}^0(0) \vert U({\vec b}) \Psi_{JM}\rangle}
{\int {\d}\vec b \, {\d}\vec b' \,
j_0\left( \vert \frac{{{\vec b}}-{{\vec b'}}}{2} \vert  q \right)
\, \langle U({\vec b'}) \Psi_{JM} \vert U({\vec b}) \Psi_{JM} \rangle}
\, .
\label{Gej}
\ee

The integration on  $\hat{\vec q}$ 
allows for further  simplifications. 
In fact,
we can now prove that the translation operator $U({\vec b})$ and the
rotation operator $R(\Omega )$, which enters the projector on angular
momentum
[Eq.  (\ref{310})], can be exchanged in the previous formula, although
they don't commute.
Let us, first of all, note that
\be
U^{^{\dag}}({\vec b}')J^0_{em}(0)U({\vec b})=
J^0_{em}({\vec b}')U({\vec b}-{\vec b}')
\label{uj0}
\ee
and expand the spherical Bessel functions $j_0$ in power series of
$\vert {\vec b}\pm {\vec b'} \vert q/2$. Notice that the integration
on the direction of the momentum transfer has eliminated the
dependence of the integrand in Eq. (\ref{Gej}) on the angle between
${\vec q}$ and ${\vec b\pm\vec b'}$.
After the introduction of the new variables
\be
{\vec z}={\vec b}-{{\vec b}}' \ \ \ ,  \ \ \ {\vec y} = {{\vec b}}' \label{zy}
\ee
we can define the scalar function
\be
F_{s,m}({\vec z})
=\int {\d}{\vec y} \ \ y^{2m+s} \left( \cos \theta_{zy} \right)^{s}
J_{em}^0 ({\vec y})
\label{fz}
\ee
where $\theta_{zy}$ is the angle between the directions $\vec z$
and $\vec y$. The l.h.s. of Eq. (\ref{Gej})
can be written as a sum of terms of the form:
\be
\int {\d}{\vec z}\,  \langle \Psi_{hh} \vert
P_{JM}
F_{s,m}({\vec z}) U( {\vec z}) P_{JM}
\vert \Psi_{hh} \rangle \, z^{2k+s} \, .
\label{sumcom}
\ee

We can now apply Eq. (\ref{theo}) to each of the previous terms
so that one of the two projectors on angular momentum in Eq. (\ref{Gej})
can be eliminated both in the numerator and in the denominator.
The electric form factor can finally be written as
\bq
G_E(q^2)&=& \frac{1}{\cal N}\, \frac{E}{M_N} \,
\int {\d}\vec a \int  {\d}\vec r  \int {\d}^3 \Omega
\ j_0(q r)
{\cal D}_{MM}^{J\,^ *}(\Omega) \nonumber \\
&\times& \, < U(-\frac{{\vec a}}{2}) \Psi_{hh} \vert
J_{em}^0({\vec r}) \vert U(\frac{{\vec a}}{2}) R(\Omega) \Psi_{hh}>
 \label{GE}
\eq
where the new variables $\vec a$ and $\vec r$
are related to the previous $\vec b$
and $\vec b'$ through
\be
{\vec b'}={\vec r}-{\vec a}/2
\ee
\be
{\vec b}={\vec r}+{\vec a}/2
\ee
and the normalization factor reads 
\be
{\cal N}=\int {\d}\vec a \, \int {\d}^3  \Omega
\, j_0(\frac{q a}{2}) \,
{\cal D}_{MM}^{J\, ^*}(\Omega)
 < \Psi_{hh} \vert
U({\vec a}) R(\Omega) \vert \Psi_{hh}>.\label{norm}
\ee

The numerator of Eq.(\ref{GE}) is the sum of three
pieces [see Eq.
(\ref{312})]: the isoscalar quark, the isovector quark and the isovector
pion contributions.

\subsection{Magnetic form factor}

From the definition of the magnetic form factor (\ref{211}) and
taking again the correspondence (\ref{313}) we can write
\be
i \, \frac{G_M(q^2)}{2 M_N} ({\vec \alpha} \times {\vec q})=
\frac{E}{M_N}
\frac{\int {\d}{\vec x}\, \langle P_{\frac{{\vec q}}{2}} \Psi_{JM} \vert
{\vec J}_{em} (0) \vert P_{-\frac{{\vec q}}{2}} \Psi_{JM}\rangle}
{\langle P_{\frac{{\vec q}}{2}} \Psi_{JM} \vert
P_{\frac{{\vec q}}{2}} \Psi_{JM} \rangle} \label{GM1}
\ee
where we have introduced ${\vec \alpha} = \langle {\vec \sigma} \rangle$.

In the case of the magnetic form factors, where the space components
of the electro-magnetic current appear, it is not possible to define
a {\it scalar} function as we did in Eq. (\ref{fz}). The projection
operators will therefore always rotate in a non-trivial way the current.
To simplify the expression of the magnetic form factor we will instead make
use of
the  relations (\ref{commut}). The current
matrix element can therefore be rewritten as
\bq
\langle \Psi _{hh} \vert P_{JM} P_{\frac{{\vec q}}{2}}
&{\vec J}_{em}&(0)P_{-\frac{{\vec q}}{2}}P_{JM}\vert \Psi_{hh}\rangle
= \langle \Psi _{hh} \vert P_{TM} P_{\frac{{\vec q}}{2}}
{\vec J}_{em} (0)P_{-\frac{{\vec q}}{2}}P_{TM}\vert \Psi_{hh}\rangle
\nonumber \\
&=& \langle \Psi_{hh} \vert P_{\frac{{\vec q}}{2}} P_{TM}
{\vec J}_{em}(0)P_{TM}P_{-\frac{{\vec q}}{2}}\vert \Psi_{hh}\rangle\, .
\eq

The current contains an isoscalar and an isovector piece.
The first one commutes with the current, while the isovector one
transforms as \cite{Fa}:
\be
P^T_{MM} \, {\vec J}^{iv}_{em\, ,0}\, P^T_{MM}=
\sum_{Q=-1}^{+1} C_Q(T,M)
{\vec J}_{em,\, Q}^{iv}P^T_{M-Q,M}
\ee
where ${\vec J}^{iv}_{em\, ,Q}$ stands for
the spherical isospin  component $Q$ of the
isovector part of the vector electromagnetic current and
\be
C_Q(T,M)=\langle 1 0 ;\, T M \vert T M \rangle
\langle 1 Q ;\, T \, M \!-\!Q \vert T M \rangle .
\ee

To extract the magnetic form factor out of Eq. (\ref{GM1})
we multiply both terms of the equation by ${\vec \alpha} \times {\vec q}$
and integrate over $\hat{\vec q}$. 
As for the electric form factors, this integration is trivial because
the Fourier transform of the matrix element depends only on $q^2$.

After a straightforward algebraic derivation on obtains the isoscalar
({\em is}) part of the magnetic form factor:
\bq
\frac{G_M^{is}(q^2)}{2 M_N}&=& \frac{1}{\cal N}
\frac{E}{M_N}
\int {\d} {\vec a} \int {\d}{\vec r} \int {\d}^3 {\Omega}
\, \frac{3 j_1(qr)}{2 qr}
{\cal D}_{MM}^{J\,^ *}(\Omega) \nonumber \\
&\times &\sum_{jk} \epsilon_{3jk}  \,
 \langle U(-\frac{{\bf a}}{2}) \Psi_{hh} \vert   \,
r_j [{\vec J}_{em}^{is}({\vec r})]_k  \,
 \vert U(\frac{{\bf a}}{2}) R(\Omega) \Psi_{hh}\rangle, \label{GMis}
\eq
where  $[{\vec J}_{em}^{is}]_k$ stands for the cartesian
$k$ component of the vector electromagnetic current (isoscalar part)
and the normalization factor ${\cal N}$ is given by Eq. (\ref{norm}),
as before.

For the isovector ({\em iv}) part of the magnetic form factor we get
\bq
\frac{G_M^{iv}(q^2)}{2 M_N} &= &\frac{1}{\cal N}\,\frac{E}{M_N}  \,
\sum_{Q} \! C_Q \!
\int {\d} {\vec a} \int {\d}{\vec r} \int {\d}^3 {\Omega} \,
\frac{3 j_1(qr)}{2 qr}
{\cal D}_{M-Q,M}^{J\,^ *}(\Omega) \nonumber \\
&\times& \sum_{jk} \epsilon_{3jk} \,
 \langle U(-\frac{\vec a}{2}) \Psi_{hh} \vert \,
r_j [{\vec J}_{em\, , Q}^{iv}({\vec r})]_k
\, \vert U(\frac{\vec a}{2}) R(\Omega) \Psi_{hh}\rangle\, .
\label{GMiv}
\eq

\section{Results and discussion}
The nucleon electric and magnetic form factors are presented in
Figs. 1-4.
The experimental data shown were taken from Refs. \cite{Exp1,Exp2,Exp3}.
We present our results in the space like region $0\le q^2 \le 0.5$ GeV$^2$.

All the results in Figures 1--4 were obtained using
 $G=0.2$ GeV (actually
$g=0.024$ GeV and $M=1.7$ GeV but, as mentioned before, the results  basically
depend on the combination $G=\sqrt{g M}$).
No parameter was fitted to reproduce the experimental form factors.

For the sake of comparison, in Figs. 1-4 we also present the
results for the form factors computed in the static approximation, i.e.
without performing the linear momentum projection.
This is  the traditional  approximation considered in
previous calculations of the form factors in the framework of
soliton models
(see Ref. \cite{sigma} for the linear sigma model and
Ref. \cite{FN94} for  the double hump version of the chiral CDM).

As it can be seen from Figs. 1-2,
the electric form factors are rather satisfactory.
One has to take into account large incertitudes in the experimental
analysis of the electric form factor of the neutron, which is
obtained from scattering on the deuteron and depends hence on the wave function
of the latter. The effect of the projection on linear momentum is
particularly relevant in the proton electric form factor.

The magnetic form factors are less satisfactory. This is probably due to the
weakness of the spin-spin interaction obtainable in this model, at least
working within the projected mean-field approximation.

As it appears from  the figures, all the computed form factors
underestimate the data for large $q^2$.
It can be interesting to note that, also studying
structure functions, one sees that in the region of large
$x=Q^2/2M\nu$, where the momentum carried by the quarks is large, the
computed quantities underestimate the data \cite{struttura}.
These problems are probably due to the
approximate treatment of translational invariance and are
therefore not so much related to the specific model
used in this paper.

The chiral CDM has now been used to compute
many different quantities. We can try to summarize the results.
The chiral CDM gives good results if problems not involving the spin
are considered. This can be seen also from the computation of the
unpolarized structure functions \cite{struttura}. Also the study
of the transition from nuclear to quark matter within this model
seems very promising, suggesting a smooth transition between the
two phases \cite{DFT95} and giving interesting results for neutron stars
\cite{stella}.

On the other hand, since
the model does not contain 
enough tensor force, it provides  poor results for
observables which involve the spin.
The magnetic form factors are therefore not totally satisfactory and
the polarized structure functions overestimate the data
\cite{struttura}, indicating that
most of the spin is carried by the quarks because the pion is very weak.
It is not yet clear whether the weakness of the pionic field 
is intrinsic to the
model
and therefore other degrees of freedom have to be considered,
or stems from the approximations used to solve the field equations.
Concerning this second possibility, there are
indications that 
a large $\Delta$-$N$ mass splitting could be obtained
using the same ingredients considered here but
allowing the scalar and the vector diquarks to have
different radii \cite{naro}.

\ack
We thank L. Caneschi for many useful discussions and for a careful
reading of the manuscript.
M.F. is indebted to T. Neuber for many
conversations. The computer codes  used in the present
work  were based on those developed by T. Neuber {\em et
al.} \cite{NF93} for the nucleon static properties.
This work was supported in part by INFN Section of Ferrara and by
the Calouste Gulbenkian Foundation (Lisbon).


%
%

%

\newpage

\begin{figure}
\centerline{\hbox{
\psfig{figure=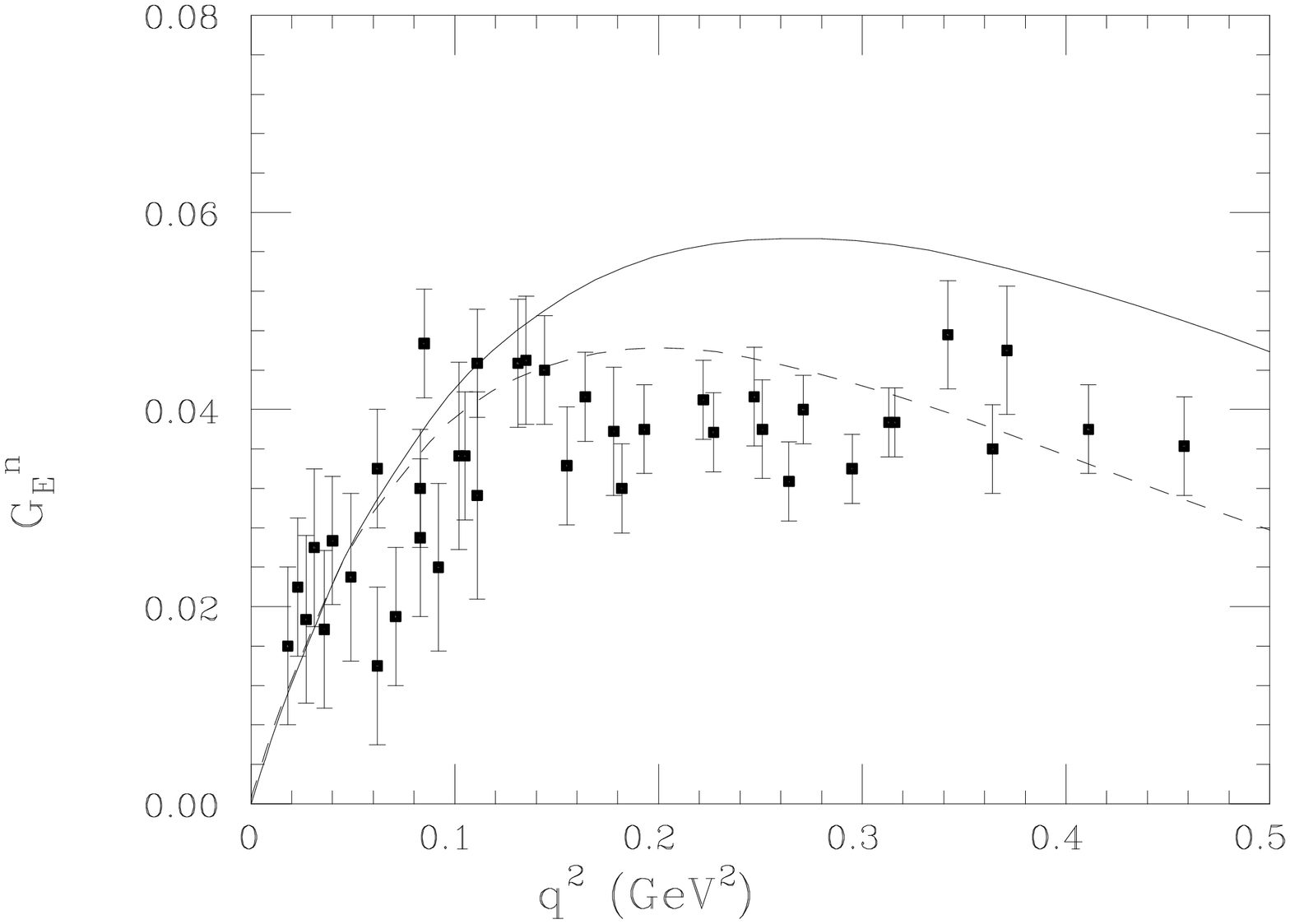,angle=90,width=14cm}
}}
{Fig. 1 \ Electric form factor of the neutron as a function
of momentum transfer in the space like region.
The dashed line corresponds to the static approximation.
The full curve corresponds to the full calculation, i.e. combined
linear and angular momentum projections.
\label{fig2}
}
\end{figure}

\begin{figure}
\centerline{\hbox{
\psfig{figure=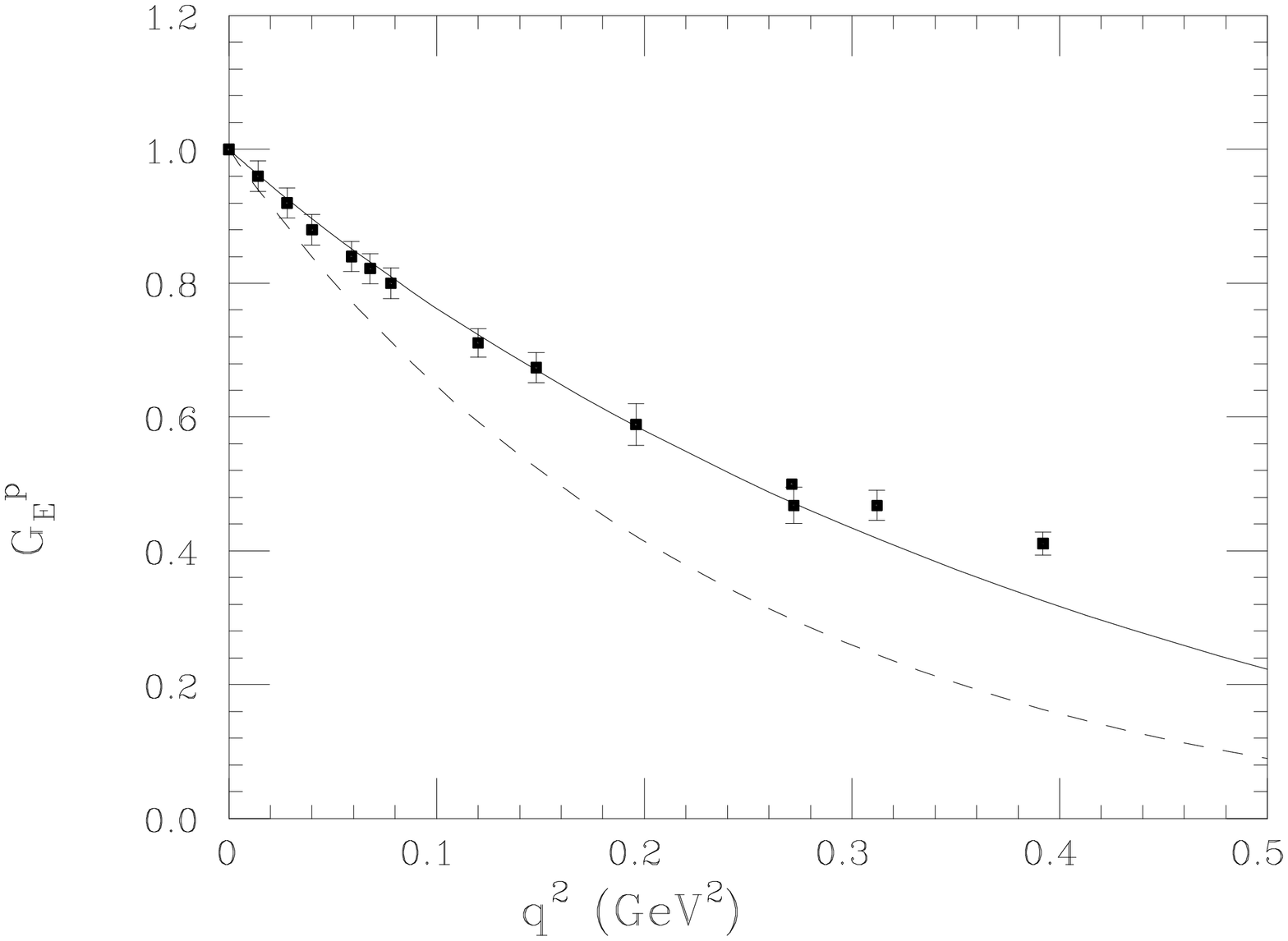,angle=90,width=14cm}
}}
{Fig. 2 \ Electric form factor of the proton. Dashed and full curves
as in Fig. 1.
\label{fig3}
}
\end{figure}

\begin{figure}
\centerline{\hbox{
\psfig{figure=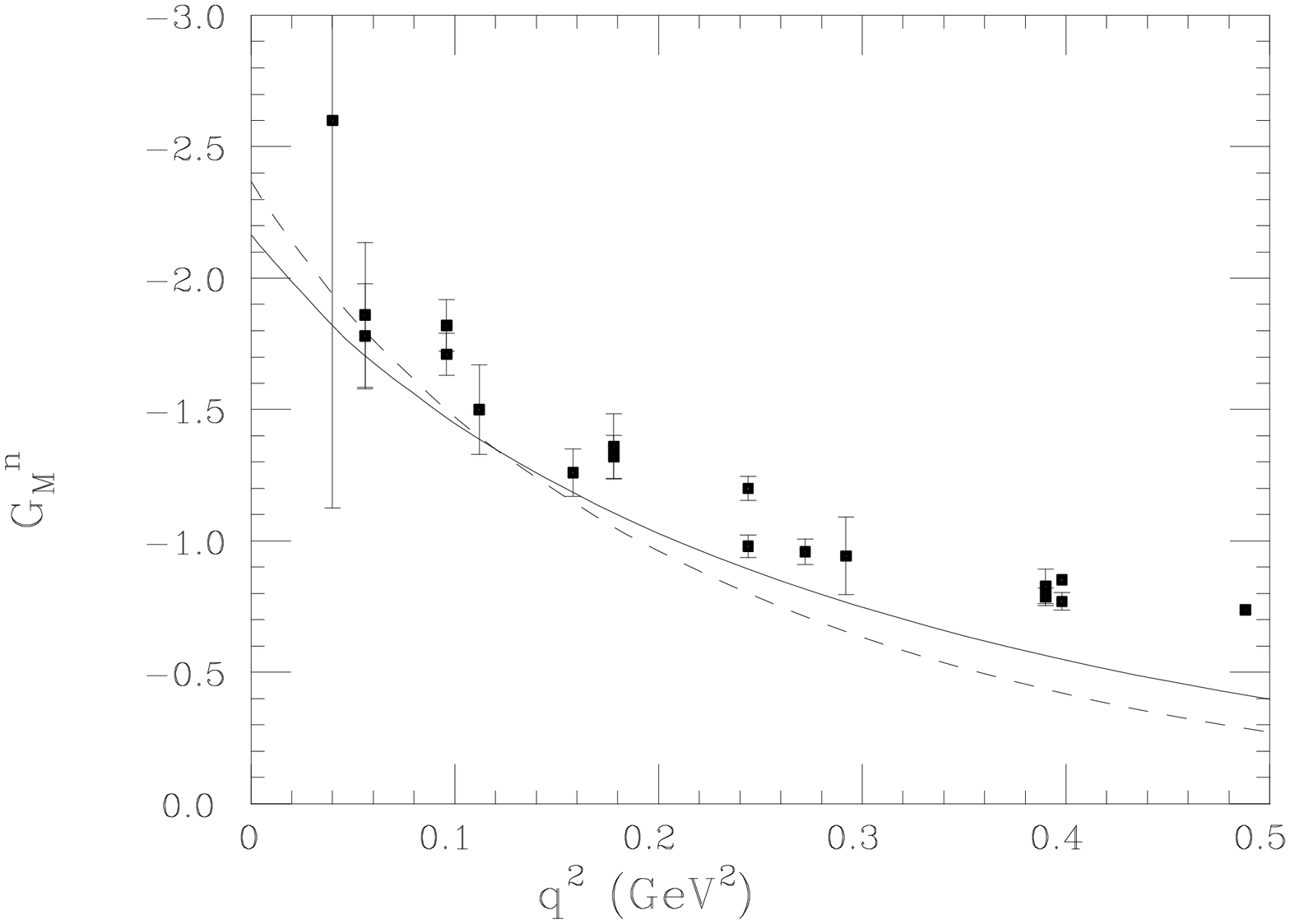,angle=90,width=14cm}
}}
{ Fig. 3 Magnetic form factor of the neutron.  Dashed and full curves
as in Fig 1.
\label{fig4}
}
\end{figure}

\begin{figure}
\centerline{\hbox{
\psfig{figure=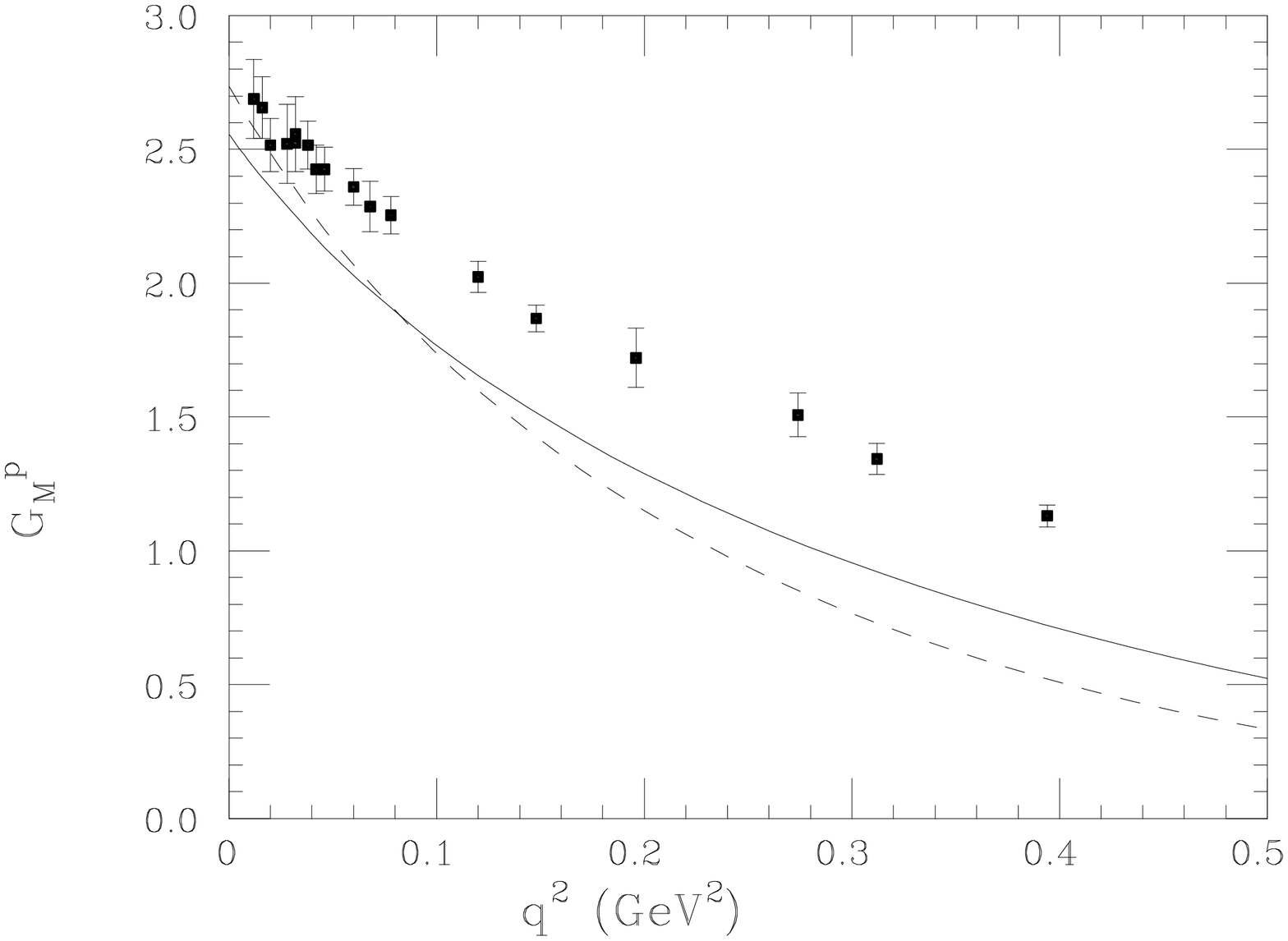,angle=90,width=14cm}
}}
{ Fig. 4 Magnetic form factor of the proton.  Dashed and full curves
as in Fig 1.
\label{fig5}  }
\end{figure}

\end{document}